\input harvmac
%\draftmode
 \noblackbox

 \newcount\figno
 \figno=0
 \def\fig#1#2#3{
 \par\begingroup\parindent=0pt\leftskip=1cm\rightskip=1cm\parindent=0pt
 \baselineskip=11pt
 \global\advance\figno by 1
 \midinsert
 \epsfxsize=#3
 \centerline{\epsfbox{#2}}
 \vskip 12pt
 {\bf Fig.\ \the\figno: } #1\par
 \endinsert\endgroup\par
 }
 \def\figlabel#1{\xdef#1{\the\figno}}
 \def\encadremath#1{\vbox{\hrule\hbox{\vrule\kern8pt\vbox{\kern8pt
 \hbox{$\displaystyle #1$}\kern8pt}
 \kern8pt\vrule}\hrule}}
 \def\apm{{\alpha^{\prime}}}

 %%% Paragraphs

 %%% special math symbols
 \font\cmss=cmss10
 \font\cmsss=cmss10 at 7pt
 \def\rlx{\relax\leavevmode}
 \def\inbar{\vrule height1.5ex width.4pt depth0pt}
 \def\IC{\relax\,\hbox{$\inbar\kern-.3em{\rm C}$}}
 \def\IN{\relax{\rm I\kern-.18em N}}
 \def\IP{\relax{\rm I\kern-.18em P}}
 \def\ZZ{\rlx\leavevmode\ifmmode\mathchoice{\hbox{\cmss Z\kern-.4em Z}}
  {\hbox{\cmss Z\kern-.4em Z}}{\lower.9pt\hbox{\cmsss Z\kern-.36em Z}}
  {\lower1.2pt\hbox{\cmsss Z\kern-.36em Z}}\else{\cmss Z\kern-.4em
  Z}\fi}
 %%% misc.
 \def\IZ{\relax\ifmmode\mathchoice
 {\hbox{\cmss Z\kern-.4em Z}}{\hbox{\cmss Z\kern-.4em Z}}
 {\lower.9pt\hbox{\cmsss Z\kern-.4em Z}}
 {\lower1.2pt\hbox{\cmsss Z\kern-.4em Z}}\else{\cmss Z\kern-.4em
 Z}\fi}
 %%% misc.
 \def\IZ{\relax\ifmmode\mathchoice
 {\hbox{\cmss Z\kern-.4em Z}}{\hbox{\cmss Z\kern-.4em Z}}
 {\lower.9pt\hbox{\cmsss Z\kern-.4em Z}}
 {\lower1.2pt\hbox{\cmsss Z\kern-.4em Z}}\else{\cmss Z\kern-.4em Z}\fi}

 \def\narrowplus{\kern -.04truein + \kern -.03truein}
 \def\narrowminus{- \kern -.04truein}
 \def\narrowminussub{\kern -.02truein - \kern -.01truein}
 
 \def\half{{1\over 2}}
 
 \def\m{{\mu}}
 \def\n{{\nu}}
 \def\ep{{\epsilon}}
 \def\d{{\delta}}
 \def\t{{\theta}}
 \def\a{{\alpha}}
 \def\T{{\Theta}}
 \def\frac#1#2{{#1\over #2}}
 \def\l{{\lambda}}
 
 \def\D{{\Delta}}
 \def\g{{\gamma}}
 \def\s{{\sigma}}
 \def\ch{{\chi}}
 \def\b{{\beta}}

 \def\p{\partial}

 \def\apm{{\a^{\prime}}}

 \def\IZ{\relax\ifmmode\mathchoice
 {\hbox{\cmss Z\kern-.4em Z}}{\hbox{\cmss Z\kern-.4em Z}}
 {\lower.9pt\hbox{\cmsss Z\kern-.4em Z}}
 {\lower1.2pt\hbox{\cmsss Z\kern-.4em Z}}\else{\cmss Z\kern-.4em Z}\fi}
 \def\IB{\relax{\rm I\kern-.18em B}}
 \def\IC{{\relax\hbox{$\inbar\kern-.3em{\rm C}$}}}
 \def\ID{\relax{\rm I\kern-.18em D}}
 \def\IE{\relax{\rm I\kern-.18em E}}
 \def\IF{\relax{\rm I\kern-.18em F}}
 \def\IG{\relax\hbox{$\inbar\kern-.3em{\rm G}$}}
 \def\IGa{\relax\hbox{${\rm I}\kern-.18em\Gamma$}}
 \def\IH{\relax{\rm I\kern-.18em H}}
 \def\II{\relax{\rm I\kern-.18em I}}
 \def\IK{\relax{\rm I\kern-.18em K}}
 \def\IP{\relax{\rm I\kern-.18em P}}
 %\def\IX{\relax{\rm X\kern-.01em X}}
 %this doesn't work

 \font\cmss=cmss10 \font\cmsss=cmss10 at 7pt
 \def\IR{\relax{\rm I\kern-.18em R}}

 %

 %
 %       \eqn\label{a+b=c}       gives displayed equation, numbered
 %                               consecutively within sections.
%     \eqnn and \eqna define labels in advance (of eqalign?)
 %
 \def\eqnn#1{\xdef #1{(\secsym\the\meqno)}\writedef{#1\leftbracket#1}%
 \global\advance\meqno by1\wrlabeL#1}
 \def\eqna#1{\xdef #1##1{\hbox{$(\secsym\the\meqno##1)$}}
 \writedef{#1\numbersign1\leftbracket#1{\numbersign1}}%
 \global\advance\meqno by1\wrlabeL{#1$\{\}$}}
 \def\eqn#1#2{\xdef #1{(\secsym\the\meqno)}\writedef{#1\leftbracket#1}%
 \global\advance\meqno by1$$#2\eqno#1\eqlabeL#1$$}

 \lref\sw{N.
 Seiberg and E. Witten, ``String Theory and Noncommutative
 Geometry'', hep-th/9912072.}
\lref\berg{E.~Bergshoeff,
 D.~S.~Berman, J.~P.~van der Schaar and P.~Sundell, ``A
 noncommutative M-theory five-brane,'' hep-th/0005026.}
\lref\ns{N.~Nekrasov and A.~Schwarz,
``Instantons on noncommutative R**4 and (2,0)
superconformal six  dimensional theory,''
Commun.\ Math.\ Phys.\  {\bf 198}, 689 (1998)
[hep-th/9802068].}
\lref\za{B.~Zwiebach,
``A solvable toy model for tachyon condensation in string field theory,''
hep-th/0008227.}
\lref\zb{
J.~A.~Minahan and B.~Zwiebach,
``Field theory models for tachyon and gauge field string dynamics,''
hep-th/0008231.}
\lref\david{
J.~R.~David,
``Tachyon condensation in the D0/D4 system,''
hep-th/0007235.}
\lref\bffs{T.~Banks, W.~Fischler, S.~H.~Shenker and L.~Susskind,
``M theory as a matrix model: A conjecture,''
Phys.\ Rev.\  {\bf D55}, 5112 (1997)
[hep-th/9610043].}
\lref\gna{D.~J.~Gross and N.~A.~Nekrasov,
``Monopoles and strings in noncommutative gauge theory,''
JHEP {\bf 0007}, 034 (2000)
[hep-th/0005204].}
\lref\gnb{D.~J.~Gross and N.~A.~Nekrasov,
 ``Dynamics of strings in noncommutative gauge theory,''
hep-th/0007204.}
\lref\vafa{C.~Vafa,
``Instantons on D-branes,''
Nucl.\ Phys.\  {\bf B463}, 435 (1996)
[hep-th/9512078].}
\lref\douglas{M.~R.~Douglas,
``Branes within branes,''
hep-th/9512077.}
\lref\rsennew{A.~Sen,
``Some issues in non-commutative tachyon condensation,''
hep-th/0009038.}
\lref\sena{A.~Sen,
``Non-BPS states and branes in string theory,''
hep-th/9904207..}
 \lref\hklm{
 J.~A.~Harvey, P.~Kraus, F.~Larsen and E.~J.~Martinec,
 ``D-branes and strings as non-commutative solitons,''
 hep-th/0005031.}
 \lref\gms{R.~Gopakumar, S.~Minwalla and A.~Strominger,
 ``Noncommutative solitons,''
 JHEP {\bf 0005}, 020 (2000)
 [hep-th/0003160].}
 \lref\wsft{E.~Witten,
 ``Noncommutative Geometry And String Field Theory,''
 Nucl.\ Phys.\  {\bf B268}, 253 (1986).}
 \lref\romans{L.~J.~Romans,
 ``Operator Approach To Purely Cubic String Field Theory,''
 Nucl.\ Phys.\  {\bf B298} (1988) 369.}
\lref\tachvac{ A.~Sen and  B. Zwiebach,
 ``Tachyon condensation in string field theory,''
 JHEP { \bf 0003} (2000) 002.}
 \lref\moe {N.~Moeller and  W.~Taylor, ``
 Level truncation and the tachyon in open bosonic string field theory,''
 hep-th/0002237 .}
 \lref\senb{ A.~Sen, ``Descent Relations Among Bosonic D-branes,''
  Int.J.Mod.Phys. { \bf A14} (1999) 4061-4078, hep-th/9902105.}
 \lref\evid{J.~A.~Harvey and P.~Kraus,
 ``D-branes as unstable lumps in bosonic open string field theory,''
 JHEP {\bf 0004}, 012 (2000)
 [hep-th/0002117].}
 \lref\jev{R.~de Mello Koch, A.~Jevicki, M.~Mihailescu and R.~Tatar,
 ``Lumps and p-branes in open string field theory,''
 Phys.\ Lett.\  {\bf B482}, 249 (2000)
 [hep-th/0003031].}
 \lref\evidb{N.~Moeller, A.~Sen and B.~Zwiebach,
 ``D-branes as tachyon lumps in string field theory,''
 hep-th/0005036.}
 \lref\rms{K.~Dasgupta, S.~Mukhi and G.~Rajesh,
 ``Noncommutative tachyons,''
 JHEP {\bf 0006}, 022 (2000)
 [hep-th/0005006].}
 \lref\senc{A.~Sen,
  ``Universality of the tachyon potential,''
 JHEP {\bf 9912} (1999) 027
 [hep-th/9911116].}
 \lref\send{A.~Sen,
 ``Supersymmetric world-volume action for non-BPS D-branes,''
 JHEP {\bf 9910}, 008 (1999)
 [hep-th/9909062].}
 \lref\gar{M. Garousi, ``Tachyon couplings on non-BPS D-branes and
 Dirac-Born-Infeld action'', hep-th/0003122.}
 \lref\berg{E.~A.~Bergshoeff, M.~de Roo, T.~C.~de Wit, E.~Eyras and S.~Panda,
 ``T-duality and actions for non-BPS D-branes,''
 JHEP {\bf 0005}, 009 (2000)
 [hep-th/0003221].}
 \lref\klu{J.~Kluson,
 ``Proposal for non-BPS D-brane action,''
 hep-th/0004106.}
 \lref\cubic{
 G.~T.~Horowitz, J.~Lykken, R.~Rohm and A.~Strominger,
 ``A Purely Cubic Action For String Field Theory,''
 Phys.\ Rev.\ Lett.\  {\bf 57}, 283 (1986).}
 \lref\shenker{S.~Shenker, ``What are Strings Made of?''
 talk given at ``String Theory at the Millenium'',
 http://quark.theory.caltech.edu/people/rahmfeld/Shenker/fs1.html}
 \lref\sig{See for example, W. Seigel,
 ``Covariantly Second Quantized String.2; 3,'' Phys.\ Lett.\  {\bf B149}, 157, 162
 (1984); {\bf B151}, 391, 396, (1984); T. Banks and M. Peskin,
 ``Gauge Invariance of String Fields''
 Nucl. Phys. {\bf B264}, 513 (1986). }
 \lref\rast{L.~Rastelli and B.~Zwiebach,
 ``Tachyon potentials, star products and universality,''
 hep-th/0006240.}
 \lref\schom{V.~Schomerus,
 ``D-branes and deformation quantization,''
 JHEP {\bf 9906}, 030 (1999)
 [hep-th/9903205].}
 \lref\kost{V.~A.~Kostelecky and S.~Samuel,
 ``The Static Tachyon Potential In The Open Bosonic String Theory,''
 Phys.\ Lett.\  {\bf B207}, 169 (1988).}
 \lref\padic{D.~Ghoshal and A.~Sen,
 ``Tachyon condensation and brane descent relations in p-adic string  theory,''
 hep-th/0003278.}
 \lref\witnew{E. Witten, ``Noncommutative Tachyons and String Field
 Theory'', hep-th/006071.}
 \lref\terashima{
 S.~Terashima,
 ``A note on superfields and noncommutative geometry,''
 Phys.\ Lett.\  {\bf B482}, 276 (2000)
 [hep-th/0002119].}
 \lref\jmw{
 D.~P.~Jatkar, G.~Mandal and S.~R.~Wadia,
 ``Nielsen-Olesen vortices in noncommutative Abelian Higgs model,''
 hep-th/0007078.}
\lref\mrey{G.~Mandal and S.~Rey,
 ``A note on D-branes of odd codimensions from noncommutative tachyons,''
 hep-th/0008214.}
\lref\zhou{C.~Zhou,
 ``Noncommutative scalar solitons at finite Theta,''
 hep-th/0007255.}
 \lref\mak{A.~S.~Gorsky, Y.~M.~Makeenko and K.~G.~Selivanov,
 ``On noncommutative vacua and noncommutative solitons,''
 hep-th/0007247.}
 \lref\soch{C.~Sochichiu,
 ``Noncommutative tachyonic solitons: Interaction with gauge field,''
 JHEP {\bf 0008}, 026 (2000)
 [hep-th/0007217].}
 \lref\baklee{D.~Bak and K.~Lee,
 ``Elongation of moving noncommutative solitons,''
 hep-th/0007107.}
 \lref\gmst{R.~Gopakumar, S.~Minwalla and A.~Strominger,
 ``Symmetry restoration and tachyon condensation in open string theory,''
hep-th/0007226.}
 \lref\mina{M.~Aganagic, C.~Popescu and J.~H.~Schwarz,
 ``Gauge-invariant and gauge-fixed D-brane actions,''
 Nucl.\ Phys.\  {\bf B495}, 99 (1997)
[hep-th/9612080].}
 \lref\rseiberg{N.~Seiberg,
 ``A note on background independence in noncommutative gauge theories,
 matrix model and tachyon condensation,''
 hep-th/0008013.}
 \lref\cds{A.~Connes, M.~R.~Douglas and A.~Schwarz,
 ``Noncommutative geometry and matrix theory: Compactification on tori,''
 JHEP {\bf 9802}, 003 (1998)
 [hep-th/9711162].}
\lref\ot{ Two branes out of zero branes}
\lref\other{Noncommutative field theory from matrix theory}
\lref\gmms{ R.~Gopakumar, J.~Maldacena, S.~Minwalla and A.~Strominger,
``S-duality and noncommutative gauge theory,''
JHEP {\bf 0006}, 036 (2000)
[hep-th/0005048].}
\lref\om{R.~Gopakumar, S.~Minwalla, N.~Seiberg and A.~Strominger,
``OM theory in diverse dimensions,''
JHEP {\bf 0008}, 008 (2000)
[hep-th/0006062].}
\lref\gns{E.~Gava, K.~S.~Narain and M.~H.~Sarmadi,
``On the bound states of p- and (p+2)-branes,''
Nucl.\ Phys.\  {\bf B504}, 214 (1997)
[hep-th/9704006].}
\lref\aki{A.~Hashimoto and W.~Taylor,
``Fluctuation spectra of tilted and intersecting 
D-branes from the  Born-Infeld action,''
Nucl.\ Phys.\  {\bf B503}, 193 (1997)
[hep-th/9703217].}
\lref\bak{D.~Bak,
``Exact multi-vortex solutions in noncommutative Abelian-Higgs theory,''
hep-th/0008204.}
\lref\gs{D.~Ghoshal and A.~Sen,
``Tachyon condensation and brane descent relations in p-adic string  theory,''
Nucl.\ Phys.\  {\bf B584}, 300 (2000)
[hep-th/0003278].}
\lref\poly{A.~P.~Polychronakos,
``Flux tube solutions in noncommutative gauge theories,''
[hep-th/0007043].}
\lref\ito{B.~Chen, H.~Itoyama, T.~Matsuo and K.~Murakami,
`p p' system with B field, branes at angles and noncommutative geometry,''
Nucl.\ Phys.\  {\bf B576}, 177 (2000)
[hep-th/9910263],
B.~Chen, H.~Itoyama, T.~Matsuo and K.~Murakami,
``Worldsheet and spacetime properties of p - p' system with B field  and noncommutative geometry,''
hep-th/0005283.}

\Title
 {\vbox{
 \baselineskip12pt
 \hbox{hep-th/0009142}\hbox{}\hbox{}
}}
 {\vbox{
 \centerline{Unstable Solitons in Noncommutative Gauge Theory}
 \centerline{}
 }}

 \centerline{ Mina ${ \rm Aganagic}$,
Rajesh ${\rm Gopakumar}$, Shiraz ${\rm Minwalla}$,
 and Andrew ${\rm Strominger}$}
 \bigskip\centerline{ Jefferson Physical Laboratory}
 \centerline{Harvard University}
\centerline{Cambridge, MA 02138}
 \smallskip

 \vskip .3in \centerline{\bf Abstract}
 { We find a class of exact solutions 
of noncommutative gauge theories corresponding to
unstable non-BPS solitons. In the two-dimensional
euclidean (or 2+1 dimensional lorentzian) 
$U(1)$ theory we find localized solutions 
carrying nonzero magnetic flux. In four
euclidean dimensions we find non-BPS solutions with the same
Pontrjagin charge but greater energy than the usual self-dual
Yang-Mills instanton.
We conjecture that these solutions and generalizations thereof correspond to 
nonsupersymmetric configurations of D-$(p-2k)$ branes (or
intersections thereof) in a D-$p$ brane in the noncommutative scaling
limit of large $B$-field. In the particular case of a 0-brane on a 2-brane the analysis
of small fluctuations reveals an infinite tower of
states which agrees exactly with that of the $0-2$ CFT in the
scaling limit. The spectrum contains a tachyon, and
we show explicitly that the endpoint of tachyon condensation corresponds
to the 0-brane dissolved in the 2-brane. }

 \smallskip \Date{}
\listtoc \writetoc

\newsec{Introduction}
The study of classical solutions of  noncommutative field theories
has led to a number of surprises and new insights. 
Instanton solutions to four-dimensional noncommutative Yang-Mills theory,
that elegantly resolve singularities in the moduli space, were 
found in \ns. Solitons in a wide class of
scalar field theories were found in \gms. 
These carry no topological charge and do
not saturate a BPS bound. Other interesting solutions have been
discussed in \refs{\gna \gnb \witnew \jmw \mrey \zhou \mak \soch \baklee -\gmst}.

In this paper we construct solitonic solutions of non-commutative
gauge theories. We find exact classical solutions which carry various kinds of
fluxes but, as in \gms,  are
not, in general, BPS (in contrast to the gauge theory solutions in
\ns). In the context of
string theory we argue that they correspond to non-supersymmetric
configurations of branes localized on other branes. In general it is 
difficult to find exact non-BPS solutions in an interacting gauge theory. 
Here it is possible using the representation of noncommutative gauge
fields as operators.

The first solution we construct is a soliton in 2+1 dimensional
noncommutative $U(1)$ gauge theory. It carries a unit of magnetic
flux and has finite localized energy. 
No such soliton exists in an ordinary (commutative) gauge theory. 
We conjecture that the
noncommutative solution 
corresponds to an undissolved 0-brane on a 2-brane. Evidence for
this conjecture is obtained from a comparison of energies as well as the
spectrum of fluctuations with the corresponding CFT.
We find a match in full detail with the $0-2$ system in the appropriate
noncommutative Yang Mills 
scaling limit.
The
spectrum of the $0-2$ strings consists, in this limit, of 
a single oscillator tower of massive states with the expected
tachyon at the bottom. We further work out the complete lagrangian for
the modes relevant to condensing the tachyon which allows us to follow
this process to its endpoint.\foot{An analysis of tachyon condensation
in string theory,
complementary to the one here in the non-commutative gauge theory, was
carried out in \gns\ in essentially the same system.} 
This is seen to
correspond to the expected configuration of a 0-brane dissolved in the
2-brane.

We also construct new solutions in 4-dimensional euclidean
noncommutative gauge theory with a single unit of Pontrjagin
charge. The solutions have the opposite Pontrjagin charge
of the solutions of \ns . They are non BPS at generic $\T$ 
but BPS when $\T$ is selfdual. We will argue  that this fits
perfectly with their interpretation as 0-branes on a 4-brane.

A variety of other solutions, corresponding to multiple, intersecting,
and higher codimension branes within branes are briefly described.

It may be useful to contrast the present work 
with the construction of D-branes 
as noncommutative tachyon solitons \refs{\rms, \hklm, \witnew, \gmst}.
Following \sena\ ,
these papers construct D-branes as tachyon lumps (in the bosonic theory)
or tachyon vortices (in the supersymmetric theory) asymptoting to the 
closed string vacuum. In this paper we 
construct branes as flux lumps within higher-dimension branes (following
\refs{\douglas,\vafa}).
In describing branes as tachyon lumps or 
vortices,  one faces the twin difficulties 
of the lack of a controlled approximation justifying the dynamics
and of understanding the enigmatic nothing state outside the brane. 
Both these difficulties are circumvented in this paper. 
The effective dynamics 
we use is systematically derived in a limit \sw\ in which the string scale 
is sent to infinity while keeping the noncommutativity scale fixed.
Further, our solitons asymptote 
to the well understood vacuum on higher dimensional branes, rather 
than the mysterious nothing state, which, 
unfortunately, 
cannot be studied in our setting. Nevertheless
the system we consider is quite rich and does contain tachyons and 
tachyon condensation.

The reader who is interested only in the classical solutions and
not their interpretation in string theory, need consult only
sections  2 and 3. 
Technical details are presented in in several appendices. As this
work was nearing completion, an interesting paper \bak\ appeared which also
presents the solution of section 2.1. 
{\it Note Added:} After the first version of this paper appeared we learnt
that the solutions in section 2.1 had been first written down by 
Polychronakos \poly.

\newsec{Classical Solutions in $2+1$ dimensional Gauge Theory}

In this section we present a set of exact localized solutions to
 the equations of motion of spatially noncommutative Yang Mills 
theory in 2+1 dimensions. In subsection 2.1 we present solutions 
for localized lumps of flux in pure Yang Mills theory.
In subsection 2.2 we explain how the soliton moduli space is enlarged 
in theories with additional scalar fields in the adjoint
(e.g. Super Yang Mills with 16 supercharges). In subsection 
2.3 we analyze the spectrum of small fluctuations about these
solitonic lumps.

\subsec{Codimension Two Flux Lumps in Pure Yang Mills} 
Consider U(1) Yang Mills
on a 2+1
dimensional noncommutative space with spatial noncommutativity
\eqn\comrel{[z,\bar z]=i\T^{z\bar z }=\t.}
We will take the metric $G_{\m\n}$ to be diagonal, and let
$G_{00}=-1$, $G_{z \bar z}=G$ (we will mostly set $G=1$ reinstating it when
necessary using coordinate invariance).\foot{We use the notations and
conventions of \gmst\ (see especially section 2.2) in what
follows.} In the temporal $A_0=0$ gauge, the Yang Mills action is 
\eqn\lagsimp{\eqalign{-S=&{1\over
4g^2_{YM}}\int d^3x \left(F_{\m\n}F^{\m\n} \right) \cr
=&{2 \pi\t \over g^2_{YM}} \int dt \Tr\Bigg[ - \p_t \bar C \p_t C
+ \left([C, \bar
C]+ {1\over \t}\right)^2\Bigg]}}
where we have set $C_z=C=-iA_z+a^\dagger$, $C_{\bar
z}=\bar C$, $[a,a^\dagger]={1 \over \t}$. 

The equations of motion for the dynamical field $C_z$ in the $A_0=0$ gauge is
\eqn\eom{{\p^2_t C= [C,[C, \bar
C]]. }}
In this gauge one must also impose the Gauss law constraint 
\eqn\glb{[\bar C, \p_t C]+[C, \p_t \bar
C]=0}
obtained by varying \lagsimp\
with respect to $A_0$ prior to gauge fixing, and then setting $A_0=0$. 

We will find a series of static localized solutions
to \eom, labeled by a positive integer corresponding to the number of units of flux they
carry. The solution with $m>0$ units of flux is
\eqn\nfa{C=C_0\equiv(S^\dagger)^m a^\dagger S^m 
+ \sum_{a=0}^{m-1} c^a |a\rangle \langle a|}
where
$c^a$ are arbitrary real numbers and the
shift operator $S=\sum_{i=0}^{\infty}|i\rangle \langle  i+1|$.
$S$ obeys \eqn\bash{\eqalign { &SS^\dagger=1, ~~~ S^\dagger S=1-P_0
\cr & S^m (S^\dagger)^m=1,(S^\dagger)^m S^m=1-P_{m-1}~~~ }} where
$P_{m-1}=\sum_{i=0}^{m-1}|i\rangle
\langle i|$ is the rank $m$ projector.
It is useful to note that\foot{ 
Here and through the rest of this paper, we will use the
convention $|i\rangle =0$ when $i$ is negative.} $S^m
|k\rangle = |k-m\rangle, ~(S^\dagger)^m |k \rangle  =
|k+m\rangle$ . 
The matrix $C_0$ takes the block form 
\eqn\mtform{C_0=\left( \matrix{c_0
& 0 \cr 0 & a^\dagger } \right).}
Here the upper left hand $m\times m$
block $c$ is a diagonal matrix $(c_0)_{ab}=c^a \d_{ab}$.

The flux operator $-iF_{z\bar z}=F_0$ evaluated on $C_0$ is given by
\eqn\fluxm{\t F_0=1+\t [C_0,\bar C_0]=1+\t (S^\dagger)^m[a^\dagger,
a]S^m= P_{m-1}.} 
For an arbitrary configuration, the normalized integral 
of the flux over the $z$ plane may
be rewritten as a trace over the operator $F$
 \eqn\intflux{c_1={1\over
2\pi}\int  F = \t \Tr F; } from \fluxm\ and \intflux\ 
$C_0$ carries $m$ units of flux.
Since $S^m|a\rangle =0=\langle a|(S^\dagger)^m$
($a=0 \ldots m-1$) \eqn\shsol{[C_0,F_0] =0,}
and $C_0$ is a static solution to the equation of motion \eom. 
Its energy is \eqn\encma{E={ 2\pi \t \over
2 g^2_{YM}}\Tr F_0^2={m \pi \over  g^2_{YM}\t}.}
Note that the generalization of \lagsimp\
to a constant background metric (of the form described at 
the beginning of section 2.1) has solitons of energy
\eqn\encm{E={m \pi \over  g^2_{YM} \sqrt{\T^2} }, ~~~\T^2=
\T^{\m\n} \T^{\a\b} G_{\m\a} G_{\n\b}.}
In section 4 we will interpret \nfa\ as the solution corresponding
to  $m$ D0 branes at positions $c^i$ on a D2 brane.

\subsec{Generalization with Adjoint Scalar Fields}

We will now generalize the solutions of the previous subsection to 
solutions of 2+1 Yang Mills interacting with $t$ transverse scalars 
$\ph^i$. Picking the temporal gauge 
\eqn\lagsimpb{\eqalign{-S=&{1\over
4g^2_{YM}}\int d^3x \left(F_{\m\n}F^{\m\n} +2\sum_{i=1}^t D_\m \ph^i D^\m\ph^i +
\sum_{i,j=1}^t [\ph^i , \ph^j] [\ph^i, \ph^j]\right) \cr=
&{2 \pi
\t \over g^2_{YM}} \int dt \Tr\Bigg[ -\p_t \bar C \p_t C
-\sum_{i=1}^t {1\over 2}\p_t \ph^i \p_t \ph^i + {1\over 2}\left([C, \bar
C]+ {1\over \t}\right)^2\cr &~~~~+\sum_{i=1}^t [C,\ph^i][\ph^i,
\bar C] + {1 \over 4}\sum_{i,j=1}^t [\ph^i , \ph^j] [\ph^j,
\ph^i]\Bigg].}}  
The counterpart of the equation of motion \eom\ is 
\eqn\eomb{\eqalign{\p^2_t C&= [C,[C, \bar
C]]+\sum_{i=1}^t[\ph^i,[C,\ph^i]] \cr \p_t^2 \ph^i&=
[C,[\ph^i,\bar C]]+[\bar C,[\ph^i, C]]+\sum_{j=1}^t
[\ph^j,[\ph^i,\ph^j]],}}
while the Gauss law constraint \glb\ becomes 
\eqn\glbc{[\bar C, \p_t C]+[C, \p_t \bar
C]+\sum_{i=1}^t[\ph^i,\p_t \ph^i]=0.} 
The solutions \nfa\ of the pure Yang Mills theory generalize to 
\eqn\nfab{\eqalign{C=C_0&=(S^\dagger)^m a^\dagger S^m 
+ \sum_{a=0}^{m-1} c^a |a\rangle \langle a|\cr
\ph_0^i&=\sum_{a=0}^{m-1}\ph^i_a |a\rangle \langle  a|}} where
$c^a$ and $\ph^i_a$ are arbitrary real numbers. 
Note that the gauge field in \nfab\ is identical to that in \nfa; 
in particular the solutions \nfab\ also carry $m$ units of flux.
\nfab\ takes the form 
\eqn\mtform{C_0=\left( \matrix{c_0
& 0 \cr 0 & a^\dagger } \right), ~~~~~\ph_0^i=\left( \matrix{\ph^i_0
& 0 \cr 0 & a^\dagger } \right).}
where $c_0$ and  $\ph^i_0$ are $m \times m$ diagonal matrices,
$(c_0)_{ab}=c^a \d_{ab}$
and $(\ph^i_0)_{ab}=\ph^i_a \d_{ab}$.
As 
\eqn\shsolb{[C_0,F] =0, ~~~[C_0,
\ph_0^i]=0, ~~~[\ph_0^i, \ph_0^j]=0,} \nfab\ solves the 
equations of motion \eom\ and its energy is independent of 
$\ph_0$ and is given by \encm. $\phi^i_a$ will be interpreted
as the transverse positions of the $m$ 0-branes in section 4.

\subsec{Small Fluctuations}

In Appendix A we have computed the spectrum of quadratic fluctuations about
the solutions \nfab. In this subsection we list the results of 
that computation. We first describe the spectrum about the special
solution $\ph^i_0=0$, $c_0^a=0$ and then explain how the spectrum 
is modified on giving $\ph^i$ a vacuum expectation value.

Let \eqn\expgf{\eqalign{C=&C_0+ \delta C \cr \ph^i=&\d \ph^i}}
and let 
\eqn\matfluc{\d C = \left( \matrix{ A & W \cr  \bar T &
D} \right), ~~~ \d \ph^i=\left( \matrix{ \ch^i & \psi^i \cr \bar
\psi^i & \g^i}\right) .}
In Appendix A we expand the potential energy in 
\lagsimpb\ to quadratic order in 
the fields $A, W, T, D, \ch^i, \psi^i, \g^i$, 
and diagonalize this quadratic form. We find that: 
\item{a.} The fields $A$ and $\ch^i$ do not appear in the 
quadratic potential and so are massless
modes in 0+1 dimensions. In section 4 these modes will be
identified with the scalar
fields living on the world volume of $m$ coincident D0-branes.
\item{b.} The potential 
for the modes $D$ and $\g^i$ is precisely that for
fluctuations of the vector and the transverse
scalars of \lagsimp\ about its vacuum. In section 4
$D$ and $\g^i$ will be identified with the modes on a D2-brane.
\item{c.} The off diagonal gauge field fluctuation modes $W$ and $T$ 
mix with each other. These fields carry $2 m \times \infty$ complex
degrees of freedom in 0+1 dimensions. 
Half of these modes are pure gauge, and are set
to zero by the Gauss Law constraint. The other half 
form a tower of states. Each energy level of the tower has $m$ complex 
fields in the fundamental of $U(m)$. The base of the tower is tachyonic
with the $m$ complex modes
$\langle a | T | m \rangle$ ($a=0 \ldots m-1$) of mass $-{1 \over \t}$.
The rest of the tower has a harmonic oscillator spectrum, with energies
\eqn\mass{m_k^2 = {(2k+1) \over 
\t},\;\;\;\; k=1,2,\ldots}
The modes of the Hermitian field $\psi^i$ also arrange themselves  
into a harmonic oscillator tower of complex fields in the fundamental 
of $U(m)$. The tower consists of states of energy 
\eqn\massb{m_k^2 = {(2k+1)\over \t
},\;\;\;\; k=0,1,\ldots}
The tower of states found here is similar to that found in \aki .
All these modes will be identified with the modes of 0-2 strings in section 
4. For later use we note that, in the more
general backgound metric described at the beginning of section 2.1, 
the fluctuation
spectrum  is obtained by replacing $\t$ in \mass, \massb\ with 
$\sqrt{ \T^2}$  where $\T^2$ is defined in \encm.

On turning on $\ph_0^1$ vev (for simplicity keeping $c_0$ and 
$\ph^i_0=0$ for $i=2\ldots t$), the spectrum described above is 
modified (see Appendix A for details):
\item{1.}
The Gauss Law constraint sets off diagonal elements in 
$\ch^1$ to zero; diagonal components of $\chi^1$ continue to be massless
physical excitations. The fields $A_{ab}$ and
$\ch^i_{ab}$ $(i=2 \ldots t)$ $(a\neq b)$ become massive, with squared mass
\eqn\addmassb{m^2(A_{ab})=m^2(\chi^{i}_{ab})=(\ph_a-\ph_b)^2.} 
Thus a $\ph_a$ vev activates the Higgs
mechanism; the off diagonal
components of $\ch^1$ are eaten by a `vector' field which
becomes massive.

\item{2.} The 2-2 sector is unaffected by $\ph^1$.

\item{3.} The $\ph_1$ vev has a simple universal effect on
the spectrum of fluctuations of all off diagonal modes ((0-2)
strings). The squared mass of every mode with index $a (a= ~1 \ldots
m-1)$ is increased by $(\ph_a)^2$. 

\newsec{More Solutions in Higher Dimensions}

 \subsec{Codimension Four Solutions} In this
subsection we describe solutions of four-dimensional euclidean 
noncommutative gauge theories which have the same Pontrjagin
charge as the usual Yang-Mills instanton, but in general
higher action (or energy). To start with we consider the U(1)
gauge theory with no additional matter fields. 

In four dimensions the two complex shifted gauge fields
$C_m$, with $m=1,2$ are  naturally viewed as operators on
the quantum mechanical Hilbert space of two particles with creation and
annihilation operators obeying
\eqn\cran{ [a^\dagger_m,a_{\bar n}]=-i\T^{-1}_{m \bar n}.}
The noncommutative field strength is
\eqn\ncfs{F_{m\bar n}=i[C_m,C_{\bar n}]-\T^{-1}_{m \bar n}.}
The gauge field equation of motion can be written
\eqn\eomt{[C^m,[C_m,C_{\bar n}]]=0.}

A solution with a single unit of Pontrjagin charge is provided by
\eqn\drf{C_m=T^\dagger a^\dagger_m T,} where $T$ obeys
\eqn\fty{TT^\dagger=1,~~~~~~T^\dagger
T=1-P_0,~~~~~~P_0T^\dagger=TP_0=0.} In this expression $P_0$ is a
projection on to the lowest radial wave functions in $both$
complex directions, $i.e.$ a four-dimensional Gaussian centered at
the origin. There are a number of gauge equivalent expressions for
$T$, we give an example in appendix B. The field strength for this
configuration is easily seen to be \eqn\fstr{ F_{m\bar
n}=-\T^{-1}_{m \bar n}P_0,} which obeys the equation of motion as
a result of \fty. 
Note that $F$ is (anti) self-dual if and only if
$\T$ is (anti) self-dual. The Pontrjagin charge and action for the
solution \drf\ are  
\eqn\pchg{p_1={1 \over 8 \pi^2}\int d^4x F\wedge F=
{1\over 2}|\sqrt{\det \T}| \T^{-1} \wedge \T^{-1}=\pm 1} 
\eqn\chg{{g_{YM}^2 \over 4 \pi^2}S= {1 \over 8 \pi^2}\int d^4x
F^{m \bar n} F_{m \bar n} =
{1 \over 2}|\sqrt{\det \T}| \T^{-1}_{m \bar
n} (\T^{-1})^{m \bar n}.} 
The RHS of \chg\ is equal to the Pontrjagin charge
(and so the BPS bound is saturated) if and only
if $\T$ is self dual or anti-self-dual. 

A solution of noncommutative 
U(N) Yang Mills with a single unit of Pontrjagin 
charge is easily constructed by embedding \drf\ into a U(1) subgroup of U(N). 

Let us first take the noncommutativity to be self-dual ($\T^-=0$)
so that the solution \drf\ is BPS, with a self-dual field strength
$F^-=0$.
This differs from the solution of Nekrasov and Schwarz 
which is an anti-self-dual instanton with selfdual noncommutativity\foot{ 
To the best of our knowledge, explicit instanton or
anti-instanton solutions for generic $\T$, or even generic
self-dual instanton solutions for $\T^-=0$, have not been found.}.
Recall that the  self-dual instanton
moduli space is a function only of $\T^-$ \sw\ 
and so, in particular, is the 
usual instanton moduli space when $\T^-=0$. The embedding of 
\drf\ into U(N) is a self-dual instanton on the singular  
locus\foot{ In the $N=4$ SYM theory \drf\ admits  
BPS deformations corresponding to turning on the transverse scalars
$\ph^i =\zeta^i|0\rangle \langle 0|$; these deformations correspond to 
moving along the Coulomb branch (taking the anti 0-brane away from the 4-brane:
recall that the anti 0-brane, 4-brane system is SUSY when $\T^-=0$).}
in the self-dual instanton moduli space, usually referred 
to as the zero size instanton singularity. 
It is interesting that  while the self-dual instanton moduli space 
is singular, our solution is smooth in the presence of noncommutativity.

For generic $\T$, the solution \drf\ is neither self-dual nor
anti-self-dual and does not saturate the BPS bound.
If $\T$ takes the 
block diagonal form $\T^{\m\n}=(\t_1 \ep, \t_2 \ep)$, 
(where $\ep$ is the $2 \times 2$ antisymmetric matrix with $\ep^{12}=1$),
the action of \drf\ exceeds that of the corresponding BPS solution
by 
\eqn\diffen{S-S_{BPS}=S-{4 \pi^2 \over g^2_{YM}}
={2 \pi^2 \over g_{YM}^2}\left( \sqrt{\t_2 \over \t_1}
-\sqrt{\t_1 \over \t_2} \right)^2.}
It is surprising
that exact, non-BPS solutions of the gauge theory can be
analytically constructed.
We conjecture that, in string theory 
in the appropriate scaling limit, \drf\ corresponds 
to an anti 0-brane sitting on top of a 4-brane. The fact that  
\drf\ is not supersymmetric is a consequence of the fact that the 
anti 0-brane 4-brane system is not supersymmetric at generic $\T$.
We will present some evidence for this conjecture in subsection 4.6.

We expect the spectrum of small fluctuations about
\drf\ at generic $\T$ to include a tachyon that becomes massless
at selfdual noncommutativity. At selfdual noncommutativity, 
(in the U(N) theory, for $N >1$), the
`tachyon hypermultiplet' potential is expected to have flat
directions, corresponding to the moduli space of dissolved self-dual 
instantons. It thus seems likely that the smooth solution \drf\ will 
permit smooth deformations into either the Higgs or the Couloumb branch
of moduli space. It would be interesting to investigate
this in more detail.

\subsec{Generalizations}

These solutions can be generalized in several ways. First, we can
translate the instanton to the position $r_m$ by adding the
term $r_m P_0$ to $C_m$ in \drf. 
Raising the $T$ and $T^\dagger$ on the right hand side of \drf\
to the $m$th power gives a solution with Pontrjagin
charge $m$ corresponding to $m$ solitons at the origin. 
The $a^\dagger$ on the right hand side of \drf\ can be
replaced by any solution of the equations of motion \eomt - for example the
general instanton solutions of \ns. Finally the construction easily
generalizes to higher dimensions. One simply finds a $T$
obeying \fty\ where $P_0$ is a higher dimensional projection operator.
Both higher codimension and intersecting brane configurations
may be described in this manner.

\newsec{D-brane Interpretation}

In this section we present evidence for the conjecture that 
the solitonic solutions constructed in previous sections represent
branes on branes. 

\subsec{The 0-2 System}

Consider a D2-brane extended in the 012 directions.
Let the closed string metric and the $B$-field in the 12 directions be
$g_{\m\n}= g \d_{\m\n}$ and $B_{\m\n}=B \ep_{\m\n}$. The low energy effective
action on such a brane is \cds\ noncommutative Yang Mills
with 16 supercharges, Yang Mills coupling $g_{YM}^2$,
open string metric $G_{\m\n}=G\d_{\m\n }$ (we will not set $G=1$ in this section)
and noncommutativity parameter $\T^{\m\n}=\t \ep^{\m\n}$ in the 12 directions
where \sw\
\eqn\paramee{ \eqalign{
\t&={1\over B}\cr
G&={(2\pi \apm B)^2 \over g} \cr
g^2_{YM}&={g_{str}
2 \pi \apm^{ \half } B \over g}. }}
In \paramee\ we have retained only the leading terms
in the limit of that is of interest to
us in this paper, namely \sw\
\eqn\limlarge{\apm B \gg g.}
\limlarge\ is the limit in which the noncommutativity length scale is
much larger than the string scale
\eqn\limlargeb{\T^2=\T^{\m\n}\T^{\a\b}G_{\m\a} G_{\n\b}=
{32 \pi^4 \apm^4 B^2 \over g^2}
\gg \apm^2 .}
In this limit, all stringy
corrections are irrelevant to processes that occur at the length and
energy scale of noncommutativity, and such processes are accurately
described in the low energy effective theory, namely noncommutative
Yang Mills.

More specifically, in the spirit of \refs{\douglas,\vafa,\sena}, a
zero brane sitting on
top of a 2-brane may be expected to be represented as a soliton in the
2-brane string field theory. It will turn out that the size of this
soliton is set by the noncommutativity scale\foot{This is
not true of every soliton in this theory. The anti D0-brane is
a string scale soliton, and so cannot be described in the low energy
effective field theory.}, and so is much larger
than the string scale in the limit \limlarge.
 Further, as this soliton carries a single unit of
flux the field
strength associated with this soliton is small in string units
$\apm^2 G^{\m\n} G^{\a\b}F_{\m\a} F_{\n\b} \sim { g^2 \over \apm^2
B^2}\ll 1$. Thus this 0-2 system and its low energy dynamics, may
be studied in detail using noncommutative Yang Mills.

We  conjecture that the solution \nfa\ in the world volume theory
of the D2-brane represents $m$ D0-branes at transverse positions
$x^i_a={2 \pi \apm \ph^i_a}$.\foot{This interpretation for the
scalar field part of \nfa\ was advanced by J. Maldacena
(unpublished).} We will now present evidence for this conjecture.

\subsec{Matching Energies} In this subsection we compare the
energy of the solution \nfa\ with the energy of a 0-brane sitting
on top of the 2-brane. We
find exact agreement.

The energy \encm\ of \nfa,   as a function of closed string parameters
is (for $m=1$)
\eqn\eqnerg{E={1 \over 2 g_{str}\sqrt{\apm}}
{g^2 \over (2\pi \apm B)^2}.}
On the other hand, consider a 0-brane sitting on top of a 2-brane
in the closed string background described 
in the previous subsection. We wish to determine $E_{bind}$, the
difference between the energy of this
configuration and one in which the 0-brane is completely dissolved in the
2-brane.
For this purpose we work in
commutative variables.  It is convenient to gauge away the bulk $B_{\m\n}$
field inducing an equal $F_{\m\n}$ field on the brane. Let the constant
value of $F$ be equal to $B$ after the 0-brane has dissolved into the
2-brane. The energy of this dissolved state is
\eqn\energytk{E= {1\over g_{str}(2 \pi)^{2}\apm^{{3 \over 2}}}
\int d^{2}x \sqrt{ \det{ (g+2 \pi \apm B)}}=
{1\over (2 \pi)^2 g_{str} \apm^{{3 \over 2}}}
\int d^{2}x \sqrt{g^2 +(2 \pi \apm B)^2}. }
In the limit of large noncommutativity \limlarge,
\energytk\ may be expanded as
\eqn\ene{E={1\over g_{str} \sqrt{\apm}}
{1\over 2 \pi} \int d^{2}x B\left( 1+ \half {g^2 \over (2 \pi \apm B)^2}
+ \ldots\right) .}
Removing a unit of D0-brane charge from the constant
value of the background $F$ field on the brane,
(${1 \over 2 \pi} \int d^2x \Delta F=-1$) lowers the energy of the 2-brane by
\eqn\le{\Delta E={1\over g_{str} \sqrt{\apm}}
\left( 1- \half {g^2 \over (2 \pi \apm B)^2}
\right).}
Thus
\eqn\neten{E_{bind}=E(D0)-\Delta E={1\over g_{str}\sqrt{\apm}}-\Delta E
={1\over2 g_{str}\sqrt{\apm}}{g^2 \over (2 \pi \apm B)^2}}
in precise agreement with the energy \eqnerg\ of the solution \nfa.

\subsec{Matching Spectra}

We have interpreted the solution \nfa\ as $m$ D0-branes sitting
outside a D2-brane that carries large D0-brane charge density. In
this subsection we will match the spectrum of fluctuations about
\nfa\ with the spectrum of the free 0-2 conformal field theory. Of
course we expect complete agreement only in the scaling limit
\limlarge; away from this limit the full 0-2 CFT has a Hagedorn
spectrum of string scale states. Rather surprisingly, however, the
CFT also has a harmonic oscillator tower of states whose
energies are parametrically small in string units.\foot{This was already
observed in \gns. That paper also computes the spectrum of the 
0-2 CFT described in this section. The computation of the 0-2 spectrum 
in appendix C was also carried out in \ito.
Closely 
related computations in the 0-4 system are also described in \sw, 
whose notations and conventions we follow.} These
states have masses proportional to ${1 \over \t}$ rather than 
${1 \over \apm}$ and survive the
scaling limit. They match perfectly with the spectrum of
fluctuations about the solution \nfa\ derived in 
subsection 2.3. In particular, the $B$ field shifts the 0-2 tachyon
mass-squared to just below zero, in such a way that it remains
finite in the scaling limit.

Let us first recapitulate from subsection 2.3 the relevant properties
of the soliton configuration corresponding to $m$  0-branes
sitting on top of the 2-brane, i.e. the solution \nfa\ with
$\ph^i_a=0$. Using \paramee, and in particular \eqn\comp{
\frac{1}{G\theta}=\frac{g}{4 \pi^2 \apm^2 B },} the fluctuation
spectrum derived in section 2.2 consists of
\item{a.} Massless $m \times m$ transverse scalars,
on the 0-branes.
\item{b.} Massless fluctuations of the gauge field and transverse scalars
on the 2-brane.
\item{c.} A single complex tachyon (in the fundamental of $U(m)$)
with mass $m^2=-{g \over 4 \pi^2 \apm^2 B}$ and 
a harmonic tower of states with $m^2=(2k+1) {g\over 4 \pi^2 \apm^2 B}
~~~k=1,2  \ldots$ (from the gauge modes). One additional such
tower with states with $m^2=(2k+1){g \over 4 \pi^2 \apm^2 B}
~~~k=0,1  \ldots$ for each transverse scalar.

We now consider the direct computation of the 0-2 spectrum from
the free conformal field theory that describes this system. Our
detailed analysis of the spectrum follows that given
for the 0-4 system in \sw. 
Here we summarize this analysis; some additional details are presented in 
Appendix C.

The spectrum of 0-0 strings is unaffected by the $B$ field. Only
the massless states are light in string units; their spectrum
is exactly as in (a.) above. Similarly the only light 2-2
states are the massless fields, and they match precisely with (b.)
above. Now consider 0-2 strings. The part of the conformal field
theory that involves the DD and NN directions (we work in the 
covariant formalism) is unaffected by the
magnetic field, and may be dealt with as usual.
The interesting effects occur in the N-D directions (directions 1 and 2). 
The $B$ field modifies boundary conditions of these strings, 
influencing their mode expansion. It turns out that, 
in the NS sector,  the fields $X=x^1+ix^2$, 
$\bar X= X^\dagger$,  $\psi =\psi^1+i \psi^2$ and $\bar \psi$ 
are mode expanded in terms of the operators 
$\a_{n+\n}$, $\bar \a_{n-\n}$, $\psi_{n+\n -\half}$ and 
$\bar \psi_{n - \n + \half}$ respectively, where $n$ runs over all 
integers and, as usual, the subscript of an oscillator field 
represents its $L_0$ eigenvalue, $[L_0, \b_r]=-r\b_r$ ($\b$ represents
any of the oscillators above). The constant $\n$ depends on the   
magnetic field, and, in the limit \limlarge,  is given by  
\eqn\lim{\nu = 1 - \frac{1}{\pi b}, ~~~~b
=\frac{ 2 \pi \apm B}{g}.}

Summing up all zero point energies, the energy of the vacuum state is \sw\ ,
\eqn\ee{E_{vac}=-\frac{1}{2}(|\nu-\frac{1}{2}|+\frac{1}{2}),}
and, at large $b$, $E_{vac}=-\half( 1-{1\over \pi b}).$ 
The state $|0\rangle$ is thus a spacetime tachyon 
with string scale mass. 
However, the GSO
projection removes from the spectrum\foot{
The opposite GSO projection that retains states with even
fermion number and projects out states with
odd fermion number yields the conformal field theory for an anti zero
brane sitting on top of the 2-brane \sw.}
$|0\rangle$ and all other states with even fermion number.
The lowest energy state retained by the GSO projection is
$\bar\psi_{-\nu+\frac{1}{2}}| 0 \rangle$ and has energy
\eqn\energy{E_{vac}+ \nu - \frac{1}{2}=-{1\over 2 \pi b}.}
This state corresponds to a spacetime tachyon. Note that its mass 
is parametrically
small in string units, due to a cancellation of the leading terms. 
Another light state
is $\psi_{-1+(\nu-\frac{1}{2})}|0\rangle $, 
the energy of the state is ${3\over 2 \pi b}$.
Other states created by ND fermion oscillators have large,  
string scale masses. 
Turning to the remaining fields in the CFT, any of the lowest
transverse (NN or DD) fermionic oscillators create states 
$\psi^{\mu}_{-\half}|0\rangle$ with small positive mass, ${1 \over 2 \pi b}$.
Finally, there exists a single bosonic oscillator
$\a_{-1+\n}$ whose energy ${1 \over \pi b}$
is parametrically small in string units. This oscillator turns each of
each of $\bar\psi_{-\nu+\frac{1}{2}}|0\rangle$, 
$\psi_{-1+(\nu-\frac{1}{2})}|0\rangle$, and $\psi^{i}_{-\half}|0 \rangle$ into
the base of a harmonic oscillator tower of light states created by
$\bar\psi_{-\nu+\frac{1}{2}} , \psi_{-1+(\nu-\frac{1}{2})}$, and $\psi^{\mu}$
acting on $|k\rangle={1\over \sqrt{k!}}\a_{-1}^k|0 \rangle$.
 
Physical states $|\ph\rangle$ are those
\item{a.} That  
obey $L_{\{m>0\}}|\phi \rangle =0 = G_{\{m>0\}} |\phi \rangle ,$
(where $L_m$, $G_m$ are modes of
the world-sheet energy momentum tensor and supersymmetry 
current, see Appendix C),
\item{b.} Modulo states that are pure gauge, i.e. are of the form
\eqn\equv{
 |\phi \rangle = L_{\{m<0\}}|\chi \rangle 
\;\;\;\;\;or \;\;\;\;\; |\phi \rangle =
  G_{\{m<0\}}|\chi \rangle ,}for some $| \ch \rangle$. 

A general coherent state 
can be expanded as 
\eqn\dqm{|\phi\rangle =\sum_{k=0}^{\infty} 
\{ W_{k} \psi_{-1+(\nu-\frac{1}{2})}+T_{k}\bar\psi_{-\nu+\frac{1}{2}} +
(\chi_{\mu})_k \psi^{\mu}_{-\half}\}|k\rangle,}
where $W_k,T_k$ and $\chi_{\mu}$ are functions of time.
The constraints (a.) and (b.) above 
are non-trivial for operators $G_{(\pm\frac{1}{2})}.$
$G_{\half}|\phi\rangle = 0$ implies that
\eqn\cnsta{\sqrt{k} W_{k-1} + \sqrt{k+1} T_{k+1}-
\partial_t(\chi_{t})_k=0.}
Acting by $G_{-\frac{1}{2}}$
on $\phi_k|k\rangle$, where $\phi_k=\phi_k(t)$, 
we conclude that
\eqn\equivb{(W_k, T_k, \ch^0_k) \sim (W_k+\sqrt{k+1} \ph_{k+1}, 
T_k+\sqrt{k}\ph_{k-1}, \ch^0_k-\p_0 \ph_k)}  
%\eqn\equivb{\eqalign{W_{k-1}&~\sim~ W_{k-1} + \sqrt{k} \phi_k\cr
%T_{k+1}&~\sim~ T_{k+1} + \sqrt{k+1} \phi_k\cr
%(\chi_t)_{k}&~\sim ~(\chi_t)_{k} - \p_t\phi_k.}}
where $\ph_k$ are arbitrary.
We can use \equivb\ to set $(\chi_t)_k=0,(k=0,1\ldots)$,
and then the stringy constraint  \cnsta\ implies 
\eqn\qt{\sqrt{k+1}T_{k+1} 
+\sqrt{k}|W_{k-1}=0,~~~~k=0,1,\ldots}
eliminating roughly one of the two 
towers\foot{Indeed the states identified as 
pure gauge in this procedure are in precise correspondence with the 
off diagonal modes of the fluctuation gauge field that were set to zero 
by the Gauss Law constraint in the fluctuation analysis of subsection
2.3. }  
All $\chi^{i}_k (i=1,\ldots)$
states are physical. 

In summary, the spectrum of light  0-2 strings contains:
\item{i.} A tachyon $T_0$ of
squared mass $-{1 \over 2 \pi \apm b}= -{g \over 4 \pi^2 \apm^2 B}$,
\item{ii.}Tower of massive states, with $m_k^2= {(2k+1)\over 2\pi\apm b} ~~~k=1,2 \ldots$  which are linear combinations 
of 0-2 Fock-space states
${ 1\over \sqrt{2k+1}}(\sqrt{k}T_{k+1} - \sqrt{k+1}W_{k-1}).$
\item{iii.} A tower of states $\chi^i_k,$
for $k=0,1\ldots$, of masses $m_k^2 = {2k+1 \over 2 \pi \apm b}$.

The spectrum of light states in 
precise agreement with the spectrum of oscillations above the
soliton. 

Finally when the zero brane is displaced from the 2 brane by a
physical distance $x^1$ (say in the 1st transverse direction), all
squared masses computed above are then increased by ${(x^1)^2
\over 4 \pi^2 \apm^2 }$. Making the usual identification $x^i =2
\pi \apm \ph^i$, this matches the shift in squared mass of
fluctuation modes of the soliton \nfa\ on giving $\ph^i$ an
expectation value.

\subsec{Interpretation in Matrix Theory}

It is well known that a D2-brane with a background magnetic field can
be constructed out of an infinite number of D0-branes \bffs; a D2-brane
in the 12 directions may be represented as a matrix configuration
\eqn\maconst{ X^1=X^1_{0}, ~~~X^2=X^2_{0}}
where $X^1_{0}$ and $X^2_{0}$ are infinite dimensional matrices
obeying
\eqn\commrel{ [X^1_{0}, X^2_{0}]=i \t.}

\nref\li{M. Li, ``Strings from IIB Matrices,''
hep-th/9612222, Nucl. Phys. {\bf B499} (1997) 149.}%
\nref\aiikkt{H.~Aoki, N.~Ishibashi, S.~Iso, H.~Kawai, Y.~Kitazawa and
T.~Tada, ``Noncommutative Yang-Mills in IIB matrix model,''
Nucl.\ Phys.\  {\bf B565} (2000) 176 [hep-th/9908141].}%
\nref\ishibashi{N.~Ishibashi, ``A relation between commutative and
noncommutative descriptions of  D-branes,'' hep-th/9909176.}%
\nref\iikk{N. Ishibashi, S. Iso, H. Kawai and Y. Kitazawa,
``Wilson Loops in Noncommutative Yang-Mills,'' hep-th/9910004.}%
\nref\barsminic{I. Bars and  D. Minic, ``Noncommutative geometry on a
discrete periodic lattice and gauge theory,'' hep-th/9910091.}%
\nref\amns{J. Ambjorn , Y. Makeenko, J. Nishimura and R.Szabo,
``Finite $N$ matrix models of noncommutative gauge theory, ''
hep-th/9911041,  JHEP { \bf 11} 029 (1999); ``Nonperturbative dynamics
of noncommutative gauge theory,''  hep-th/0002158;
J.~Ambjorn, K.~N.~Anagnostopoulos, W.~Bietenholz, T.~Hotta and
J.~Nishimura,
``Large N dynamics of dimensionally reduced 4D SU(N) super Yang-Mills
theory,'' JHEP {\bf 0007} (2000) 013
[hep-th/0003208];
J.~Ambjorn, Y.~M.~Makeenko, J.~Nishimura and R.~J.~Szabo,
``Lattice gauge fields and discrete noncommutative Yang-Mills theory,''
JHEP {\bf 0005} (2000) 023
[hep-th/0004147].}%
\nref\agwa{L.~Alvarez-Gaume and S.~R.~Wadia, ``Gauge theory on a
quantum phase space,'' hep-th/0006219.}%
\nref\fatoll{A.~H.~Fatollahi,
``Gauge symmetry as symmetry of matrix coordinates,''
hep-th/0007023.}%

Fluctuations about this soliton describe a noncommutative gauge theory
\refs{\li-\fatoll,  \rseiberg}.
On the other hand matrices corresponding to $m$ 0-branes at specific
spatial postions are simply diagonal $m \times m $ blocks, whose
eigenvalues give the postions of each of these 0-branes.
Appending such an $m \times m$ block to the top left corner of the
matrix representing a D2-brane (and setting all off diagonal elements
to zero) yields a system consisting of $m$ D0-branes plus a D2-brane.
This configuration is clearly a solution to the equations of
motion; it would be natural to expect it to represent the soliton
in the noncommutative field theory that corresponds to $m$ D0-branes
sitting on the D2-brane.

Indeed $C_0$ (see \mtform) is precisely of this form. We regard
this observation as providing further evidence, from a different
viewpoint, for our interpretation of the soliton \nfa.\foot{We
thank Per Kraus for discussions on this
connection.}

\subsec{The D-String}

\nfa\ may trivially be
lifted to a $p-2$ dimensional soliton of $p+1$ dimensional noncommutative
Yang Mills with spatial noncommutativity in at least two directions.
The solution may then be interpreted as $m$ $D(p-2)$ branes sitting
outside a Dp-brane.

The case $p=3$ is of particular interest. In this case \nfa\
is a string like soliton in 3+1 dimensional noncommutative Yang Mills, 
and may be interpreted as a D-string parallel to the world-volume of 
a D3-brane. This is to be contrasted with the 
Gross-Nekrasov string soliton \refs{\gna, \gnb}
which may be interpreted as a D-string piercing (or ending on) the 
D3-brane at an angle. We note that the tension \encm\ of 
our non-BPS solution \nfa\ is exactly half of that of the BPS
Gross-Nekrasov string. 

3+1 dimensional Yang Mills is known to be S-dual to 3+1 dimensional
NCOS theory \gmms. Under this  S-duality, the string soliton \nfa\
maps into the NCOS string. It is easy to verify that 
the tension of the NCOS string at weak NCOS coupling
(Eqn. 2.11 of \gmms) matches its tension at strong NCOS coupling 
(Eqn. \encm\ of this paper), 
suggesting non renormalization of the NCOS string tension.

\subsec{The 0-4 System}

In Section 3.1 we conjectured that the solution \drf\ represents
an anti 0-brane sitting on top of a 4-brane in the appropriate
$B$ field. 
In this subsection we will compare the binding energy of  
\drf\ with that of an anti 0-brane sitting on top of a 4-brane 
in a large $B$ field. The two binding energies match, providing 
evidence for our conjecture. 

For simplicity we restrict attention to a closed string background 
with block diagonal $B_{\m\n}$ and $g_{\m\n}$ moduli. Let the 
$B$ field and metric along the directions of the brane be given by
$g_{\m\n}=(g_1 I_2, g_2I_2)$ and $B_{\m\n}=(B_1 \ep, B_2 \ep)$,
respectively. 
In the limit of large noncommutativity 
\eqn\limlargeb{\apm B_i \gg g_i} 
the Yang Mills coupling $g_{YM}^2$, 
open string metric $G_{\m\n}=(G_1 I_2, G_2 I_2)$ 
and noncommutativity parameter $\T^{\m\n}=(\t_1 \ep, \t_2 \ep)$ 
on the 4-brane are \sw\
\eqn\parameee{ \eqalign{
\t_i&={1\over B_i}\cr
G_i&={(2\pi \apm B_i)^2 \over g_i} \cr
g^2_{YM}&=g_{str}(2\pi)^{4} \apm^{ {5 \over 2}}
{B_1 B_2 \over g_1 g_2} . }}

The binding energy \diffen\ of the solution \drf, written as 
a function of closed string moduli is 
\eqn\bindener{E_{bind}={1\over 8 \pi^2 \apm^{{5 \over 2}}}\left[
{g_1 \over B_1}-{g_2 \over B_2} \right]^2.}
On the other hand, the binding energy of the D0-D4 system is   
the energy of a 0-brane plus the energy of a 4-brane 
minus the energy of the 0-4 bound state, all in the appropriate
background $B$ field. 
In the limit \limlarge\ it is also given by\bindener, see
\david.\foot{ See Eq. 4.10 of that paper. The translation of 
notation is as follows. $2 \pi^2 g^2$ in \david\ 
should be identified $g_{str}$ in this paper , while $b_i$ in \david\
is to be identified with  $2\pi \apm B_i$ in this paper.}

\newsec{Tachyon Condensation}

The condensation of the open string tachyon on a D-brane in
bosonic string theory and on a $D$-$\bar D$ system in type II
theory has received much recent attention (see, for instance,
\rsennew\ and references therein). One of the difficulties
encountered in these discussions is the lack of a 
parameter controlling the approximation leading to the tachyon
lagrangian. The 0-2 system we have studied in this paper also has
a world-volume tachyon, and can be regarded as a toy laboratory 
\foot{
See \refs{\gs, \za, \zb} for other models which capture some aspects of
tachyon condensation.}
for the more difficult and interesting $D$ $\bar D$ system. In the 0-2
context there is a small parameter, namely the ratio of the string scale
to the noncommutativity scale, which can be used to control the
approximations. In the scaling limit in which this ratio goes to
zero, the theory contains a tower of positive mass states as well
as the tachyon. Using the construction of the 0-brane as a soliton
\nfa\ in noncommutative Yang Mills, the lagrangian can be exactly
constructed and  tachyon condensation reliably studied.

\subsec{General Considerations}

We have argued that $m$ coincident D0-brane on a 2-brane may be
represented as the soliton \nfab\ (with $\ph^i_0=0$).
The gauge field $C$ for this solution is
\eqn\mtform{C_0=\left( \matrix{0 & 0 \cr 0 & a^\dagger } \right).}
We parameterize the arbitrary $C$ field by the fluctuation fields
of subsection 2.3
\eqn\arbcmat{C=\left( \matrix{A & W \cr \bar T & a^\dagger +D } \right).}
The fields $A$, $D$ and $W, T$ are identified with 0-0, 2-2, and 0-2
strings in the conformal field theory description of the 0-2 system.

The soliton \nfa\ is unstable to the spreading out of flux;
at the endpoint of tachyon condensation the soliton flux is spread evenly over
the infinite 2-brane and is consequently  invisible. The
final configuration is  $C=a^\dagger$. \foot{Note that 0-0, 0-2
and 2-2 fields are all excited at the endpoint of tachyon
condensation and at that point (see Appendix A for notation)
$$A=P_{m-1} a^\dagger P_{m-1},
~~~ D=S^m a^\dagger (S^\dagger)^m- a^\dagger
~~~ W=0, ~~~T=0 ~~~{\rm except}~~~ T_{m-1, 0}=\sqrt{m}.$$}
About this configuration it is most natural to parameterize 
$C= a^\dagger +A'$ where $A'$ is a new field.
The endpoint of tachyon condensation is described by the CFT
of a single 2-brane, $A'$ is identified
with the 2-2 gauge boson of this CFT.

Note that the 2-2 string modes in the CFT after tachyon
condensation include all the 0-0, 0-2, 2-0, and 2-2 strings of the
CFT prior to tachyon condensation. Thus, in the process of tachyon
condensation, 0-0 and 0-2 modes are absorbed into the 2-2
continuum. In this respect tachyon condensation in the 0-2 system
appears qualitatively different from tachyon condensation in a $p
~\bar p$ system. In the latter case there appears to be no
continuum for the $p -\bar p$ modes to disappear into. Restated,
the decay of the 0-brane into `nothing' in the 0-2 system is not
mysterious once the 0-brane is constructed as a soliton on the
2-brane.

\subsec{The Tachyon Potential}

We now study the scalar potential in this 0-2 system\foot{The
tachyon potential in the 0-4 system in a $B$ field has recently
been studied in \david.}. For notational convenience we set $\t$
to unity through the bulk of this sub-section.

The initial state $C=(S^\dagger)^m a^\dagger
S^m$ decays to the final state $C= a^\dagger$ on exciting the tachyonic mode
$T=C_{m,m-1}$. Note that the tachyonic mode and the nonzero
matrix elements in the initial and final state are all of the form
$C_{i+1,i}$. One might thus suspect that it is possible to set all
$C$ matrix elements
not of this form to zero through the entire process of tachyon
condensation.
This is indeed the case\foot{In order
to demonstrate this we assign the the fields
$C_{ij}$ and $C^*_{km}$ `angular momentum' quantum numbers
$i-j$ and $m-k$ respectively. With this assignment the potential
conserves angular momentum. All terms in the potential are
the product of an equal number of $C$ and a $C^*$ fields.
Angular momentum conservation prohibits linear coupling of `other
fields' to $C$ ( $C^*$) fields of angular momentum 1 ($-1$).}, as \lagsimp\
admits a consistent truncation to these modes.

For clarity we specialize to $m=1$. The potential expanded about
the solitonic solution \nfa\ may be written
 as a function of the tachyon
$T$ and the 2-2 gauge fields $D_{i,i-1}$ (using the notation of Appendix A).
\eqn\potlb{\eqalign{ V = {\pi \over g_{YM}^2}\Bigg[
 &\left[|T|^2-1\right]^2 +\left[|T|^2 - |D_{1,0} +1|^2 +1\right]^2 \cr & +
\sum_{i=1} \left[|D_{i,i-1} + \sqrt{i}|^2 - |D_{i+1,i} +\sqrt{i+1}
|^2 +1\right]^2 \Bigg].}}
An unstable extremum of \potlb\ that corresponds to 
an undissolved 0-brane on the 2-brane is
\eqn\frg{T=D_{i,i-1}=0.} 
It decays into the 
stable extremum \eqn\stb{|T|=1, ~~~~D_{i,i-1}=\sqrt{i+1}-\sqrt{i}.} 
It is easy to see that \stb\ corresponds to $C=a^\dagger$, the 
2-brane vacuum. 

It is possible to integrate out the fields $D_{i, i-1}$ and
obtain the potential $V$ as a function of the tachyon alone.
Minimizing $V$ w.r.t $D_{i, i-1}$ we find 
\eqn\dexp{|D_{i, i-1}+ \sqrt{i}|^2=|T|^2+i}
which sets all except the first term 
\potlb\ to zero. Restoring $\t$, the potential thus takes the
simple form\foot{ A quartic tachyon potential was also obtained in 
\gns, (see equation 2.8) using scattering calculations in string theory, 
strengthening our identification of the fluctuation modes of 
subsection 2.3 with 0-2 strings.}
 \eqn\finpot{V={ \pi \t \over g^2_{YM}}
\left[|T|^2-{1 \over \t}\right]^2.}
To be precise, in obtaining \finpot\ we have chosen the  
branch of the solution $D_{i, i+1}(T)$ that minimizes rather than
extremizes \potlb. An infinite number of other branches exist 
(and lead to the $m$ D0-brane solutions). \finpot\ is the appropriate
branch to examine the decay of the single D0-brane into the vacuum. 
{} On this branch the gauge field $C$ corresponding to a static $T$\foot{
$T$ is held static by an external source.} is 
\eqn\gfsol{C=\sum_{n=0}^\infty|n+1\rangle \langle n|\sqrt{|T|^2+n}}
and 
\eqn\fs{F=(1 - |T|^2)|0\rangle \langle 0 |.}
If we allow the tachyon to roll from $T=0$ to any particular $T$ and 
hold it at that value the localized component of the flux, it
is given by \fs\ once the 
dust has settled. 
Thus as the tachyon rolls from $|T|=0$ to the minimum at 
$|T|=1$, it radiates flux away to infinity. At $|T|=1$, the endpoint 
of tachyon condensation all the flux has been radiated away  
and the D0-brane has dissolved completely into the D2-brane.

In the case of the $D2-D4$ system, there is a $2+1$ dimensional 
complex tachyon on the world volume of the D2 brane. 
It is tempting to speculate that there are finite
energy solutions with  
one unit of magnetic flux and winding of the tachyon
which correspond to a $D0-D4$ system.

\centerline{\bf Acknowledgements}
We are grateful to J. David, 
J. Harvey, F. Larsen, E. Martinec, A. Rajaraman,
M. Rangamani, A. Sen, J. Troost and especially P. Kraus and J. Maldacena
for useful discussions.
This work was supported in part by DOE grant DE-FG02-91ER40654, NSF
grants DMS 9709694, PHY 9802709 and the Harvard Society of Fellows.

\appendix{A}{The Small Fluctuation Analysis}

We examine quadratic fluctuations about the charge $m$ solution
\nfa. For simplicity we will first consider the special case
$\ph^i_0=0$, $c^a=0$.

\subsec{$\ph^i=0$}
For convenience we set $\t=1$ in the intermediate
stages of the computation. $\t$ is restored in the final answer by
dimensional analysis.

Let \eqn\expgf{\eqalign{C=&C_0+ \delta C \cr \ph^i=&\d \ph^i.}} As
$P_{m-1} C_0 = 0 = C_0P_{m-1}$,  it is convenient to decompose the
fluctuations as \eqn\flc{\eqalign{\delta C=& A+ W+\bar T
+(S^\dagger)^m D S^m \cr \delta \ph^i=&\ch^i +\psi^i +\bar \psi^i
+(S^\dagger)^m \g^iS^m}} where \eqn\fld{\eqalign{A =& P_{m-1}
\delta C P_{m-1}, \;\;\;\;\;W = P_{m-1} \delta C (1-P_{m-1}),\cr
\bar T = &(1-P_{m-1}) \delta C P_{m-1}, \;\;\;\;\; (S^\dagger)^m
 D S^m =(1-P_{m-1}) \delta C(1- P_{m-1})}}
and \eqn\fldb{\eqalign{&\ch^i = P_{m-1} \delta \ph^i P_{m-1},
\;\;\;\;\; \psi^i= P_{m-1} \delta \ph^i(1-P_{m-1}),\cr
&(S^\dagger)^m
 \g^i S^m = (1-P_{m-1}) \delta \ph^i(1- P_{m-1}).}}

To quadratic order in fluctuations, the potential energy in
\lagsimpb\ is a linear combination of ${1 \over 2}\left([C,\bar C] +{1\over
\theta}\right)^2$ and $Tr [C,
\ph^i][\ph^i,\bar C]$, which may be  expanded as
\eqn\quadum{
{1\over 2}\Tr\left([C,\bar C]  +{1\over
\theta}\right)^2 = {m\over 2\t^2} + \Tr(W\bar C_0 - T C_0)(C_0\bar W - \bar
C_0 \bar T) + \Tr(W \bar W - \bar T T) + {1\over 2}\Tr([a^\dagger,\bar
D]+[D,a])^2,} 
\eqn\quadsc{\Tr [C_0, \ph^i][\ph^i ,
\bar C_0] =\Tr\left( (C_0 \bar C_0 + \bar C_0 C_0 ) \bar \psi^i \psi^i 
+[a^\dagger, \g^i][\g^i, a]\right).}

The fields $A$ and $\ch^i$ do not appear in \quadum\ and
\quadsc, and so are massless. 
The quadratic potential 
for the modes $D$ and $\g^i$ is precisely that for
fluctuations of the vector and the transverse
scalars of \lagsimp\ about its vacuum. Now turn to the fluctuations
in $\psi^i$. 
Expanding $\psi^i$ \eqn\exppsi{\psi^i= \sum_{k=0}^\infty
\sum_{a=0}^{m-1} \psi^i_{ak} |a\rangle\langle k+m|,} the relevant
term in \quadsc\ may be rewritten as
\eqn\quadscaa{\sum_{k=0}^\infty (2k+1) |\psi^i_{ak}|^2} {}From
\lagsimpb\ and restoring units, we conclude that the off diagonal
fluctuations of each scalar field lead to a harmonic oscillator
tower of states of mass \eqn\massb{m_k^2 = (2k+1) { 1 \over \t
},\;\;\;\; k=0,1,\ldots}

We now determine the spectrum for the off diagonal gauge field modes
$W$ and $T$. 
Let \eqn\fdef{W_{ak} =\langle a |W|k+m
\rangle , ~~~T_{ak} =\langle a |T|k+m \rangle ,} where
$a=0,...m-1$ and $k=0,...\infty$. The relevant terms in \quadum\
take the form (suppressing the summation over $a$) \eqn\quadT{
\sum_{k=0}^\infty (|\sqrt{k}W_{a,k-1}-\sqrt{k+1}T_{a,k+1}|^2) +
\sum_{k=0}^\infty(|W_{a,k}|^2-|T_{a,k}|^2)} which may be regrouped
as \eqn\massT{- |T_{a,0}|^2+ 
\sum_{k=1}^\infty(2k+1)|Y_{a,k}|^2,} where we have defined the
normalized fields \eqn\nm{Y_{a,k}=\frac{1}{\sqrt{2k+1}}
(\sqrt{k+1}W_{a,k-1}-\sqrt{k}T_{a,k+1}).} Restoring units, and
using \lagsimpb, we conclude that the spectrum of off diagonal
gauge field fluctuations consists of a single complex tachyon of
mass $-{ 1 \over \t}$, a harmonic oscillator tower of complex
modes with masses \eqn\mass{m_k^2 = (2k+1) {1\over 
\t},\;\;\;\; k=1,2,\ldots} and an infinite number of zero modes
corresponding to the orthogonal linear combinations $Z_{a,k}$
\eqn\nort{Z_{a,k}=\frac{1}{\sqrt{2k+1}}(\sqrt{k}W_{a,k-1}+\sqrt{k+1}
T_{a,k+1})~~~k=0,1 \ldots .} In fact,  the $Z_{a,k}$ fields are not
dynamical as a consequence of the Gauss law constraint \glbc.
Gauss' law, \glbc\ implies that 
to first order in fluctuations $\d C$ and $\d \ph$ about
a time independent configuration $C, \ph$ 
\eqn\glba{\p_t\left( [\bar C, \d C]+ [C , \d \bar C]+
\sum_{i=1}^t[\ph^i, \d \ph^i] \right) =0.}
Substituting $\d C=W +\bar T$ into \glba\ and using 
\fdef\ yields \eqn\elimz{\p_t\left[ Z_{ak}|a\rangle
\langle k+m| -Z_{ak}^*|k+m\rangle \langle a| \right]=0,} implying
that $Z_{ak}$ are constant in time. Thus $Z_{ak}$
are integration constants and not degrees of freedom 
and can be gauged to zero\foot{ In order to
make this completely manifest we expand the arbitrary off diagonal
gauge fluctuation in terms of $T_{a,0}$, $Y_{a,k}$ and $Z_{a,k}$
\eqn\exparboffd{\eqalign {& T_{a,o}^*|m\rangle \langle a| +
\sum_{a=0}^{m-1}\Bigg[ \sum_{k=1}^\infty{1 \over
\sqrt{2k+1}}\left( \sqrt{k+1}Y_{ak}|a\rangle \langle k+m-1|-
\sqrt{k}Y^*_{ak}|k+m+1\rangle \langle a| \right) \cr
&+\sum_{k=0}^\infty{1\over
\sqrt{2k+1}}\left(\sqrt{k}Z_{ak}|a\rangle \langle k+m-1|+
\sqrt{k+1}Z^*_{ak}|k+m+1\rangle \langle a| \right) \Bigg],}} and
note that all terms involving $Z_{ak}$ in \exparboffd\ are of the
form ${[\l, C_0]}$ where $\l$ is the anti-hermitian operator
\eqn\deflam{\l={1\over \sqrt{2k+1}}
\sum_{k=0}^{\infty}\sum_{a=0}^{m-1} \left( Z_{ak}|a\rangle \langle
k+m|- Z^*_{ak}|k+m\rangle \langle a| \right) .} Thus $Z_{ak}$ can
be set to zero at $t=0$ by a gauge transformation, and the Gauss
Law constraint ensures that $Z_{ak}$ stays zero at all times.}.

\subsec{$\ph^1 \neq 0$}

In the rest of this Appendix we will describe how the spectrum
of fluctuations is altered upon turning on a vev for $\ph_1$ in
the solution \nfab, for simplicity leaving $c^a=0$, 
$\ph^i=0 $ ($i=2, \ldots
t$). Let \eqn\phform{\ph^1=\sum_{a=0}^{m-1}\ph_a|a\rangle \langle
a |} where $\ph_a$ are arbitrary nonzero real numbers. The
analysis proceeds as above, with the following modifications.

First consider fluctuations of the fields $A_{ab}$ 
and $\ch_{ab}$. 
The Gauss Law constraint \glba\ implies \eqn\diagrest{\p_t \left(
[\ch^1, \ph^1] \right)=0} which (assuming $\ph_a \neq \ph_b$ for
$a \neq b$) implies that physical fluctuations of $\ch^1$ are
diagonal. On the other hand the remaining fields $A_{ab}$ and
$\ch^i_{ab}$ $(i=2 \ldots t)$ are now massive, with squared mass
(from the fourth and fifth terms in \lagsimpb\ respectively)
\eqn\addmassb{m^2(A_{ab})=m^2(\chi^i_{ab})=(\ph_a-\ph_b)^2.} 

The fields $D$ and $\g^i$ are unaffected by the vev for $\ph^1$.
We now turn to the off diagonal fields $T$, $W$, and $\psi^i$. 
The potential energy for off diagonal
fluctuations receives  several additions 
from the $\ph^1$ vev. {}From
the fifth term in \lagsimpb\ we find the additional contribution to
the potential \eqn\additoo{\sum_{i=2}^t \sum_{a=0}^{m-1}
(\ph_a)^2 \sum_{k=0}^\infty |\psi^i_{ak}|^2.} Thus the off
diagonal fluctuations of the scalars $\ph^i$ ($i=2 \ldots t$) pick
up an additional mass \eqn\incmassa{\D m^2(\psi^i_{ak})=  \ph_a^2.}
We now consider the gauge fields and $\psi^1$. In addition to the
potential (computed above) at $\ph_a=0$ \eqn\potsum{V=
-|T_{a,0}|^2+ \sum_{k=1}^\infty(2k+1)|Y_{a,k}|^2 
+ \sum_{k=0}^\infty (2k+1) |\psi^1_{ak}|^2 , } at nonzero $\ph_a$
we find the additional contributions (from the fourth term in
\lagsimpb) \eqn\ado{\eqalign{&\sum_{a=0}^{m-1}\sum_{k=0}^\infty
\left( \sqrt{2k+1} \ph_a\left(\psi^1_{ak}Z^*_{ak}
+(\psi^1_{ak})^*Z_{ak} \right) +
  \ph_a^2
\left(|W_{ak}|^2+|T_{ak}|^2 \right) \right) \cr =&
\sum_{a=0}^{m-1} \left[ \ph_a^2 |T_{a0}|^2 + \sum_{k=0}^\infty
\left( \sqrt{2k+1} \ph_a\left(\psi^1_{ak}Z^*_{ak}
+(\psi^1_{ak})^*Z_{ak} \right) \right) + \ph_a^2         
\left( \sum_{k=1}^\infty |Y_{ak}|^2+\sum_{k=0}^\infty 
|Z_{ak}|^2 \right)  \right].}} 

The sum of
the potentials in \potsum\ and \ado\ may be rewritten as
\eqn\netpot{V=  \sum_{a=0}^{m-1}\left[(-1 + \ph_a^2) |T_{a,0}|^2+
\sum_{k=1}^{\infty}\left( {2k+1} +(\ph_a)^2 \right)|Y_{ak}|^2 +
\sum_{k=0}^\infty ({2k+1} + (\ph_a)^2)|V_{ak}|^2 \right]} where
\eqn\defv{V_{ak}={ 1\over \sqrt{ 2k+1+ (\ph_a)^2}} \left(
\sqrt{2k+1} \psi^1_{ak} + \ph_a Z_{ak}\right).} Thus the spectrum
of 0-2 states at $\ph_a \neq 0$ is exactly that at $\ph_a=0$,
shifted by $(\ph_a)^2$. The existence of zero modes
\eqn\smu{U_{ak}={ 1\over \sqrt{ 2k+1+ (\ph_a)^2}} \left(
\sqrt{2k+1}Z_{ak}  - \ph_a \psi^1_{ak} \right)} in the fluctuation
spectrum is a consequence of the Gauss Law constraint \glba\
which, in the 0-2 sector yields \eqn\glim{\p_t \left[U_{ak}
|a\rangle \langle k+m| + U^*_{ak}|k+m\rangle \langle a|
\right]=0.} Thus $U_{ak}$ is unphysical and can be set to zero by
a choice of gauge.

\appendix{B}{The Matrix T}

In this appendix we construct a matrix $T$ obeying 
\eqn\fty{TT^\dagger=1,~~~~~~T^\dagger
T=1-P_0,~~~~~~P_0T^\dagger=TP_0=0,}
where $P_0$ is the projector onto the lowest radial wave function of 
the two-dimensional harmonic oscillator. 
It is convenient to start with an angular momentum basis
\eqn\amb{|j,m\rangle= {(a_1^\dagger)^{j+m}(a_2^\dagger)^{j-m}
\over \sqrt{(j+m)!(j-m)!}}|0,0 \rangle .}
An integer ordering for this basis can be defined for example by 
\eqn\tvc{|j^2+j+m \rangle \equiv |j,m \rangle .}
In this basis the solution for $T$ is 
\eqn\rst{ \langle k |T |l \rangle =\delta_{k,l+1}.}
This solution is of course highly non-unique, but since 
different choices of $T$ obeying \fty\ lead to the same field strength, 
one expects that they are related by $U(\infty)$ transformations, as indeed 
can be verified. 

\appendix{C}{Details of the 0-2 Spectrum}

In this Appendix we present some details omitted in
the discussion of the 0-2 CFT in subsection 4.3. This computation was also
carried out in \ito.

Let the string world-sheet be a strip,
parameterized by coordinates $0\leq \s \leq \pi$ and $-\infty < \tau <\infty$.
The $x^{1,2}$ fields on a
string beginning at the 0-brane and ending at the 2-brane obey the usual boundary
conditions $\p_{\tau} x^a|_{\s=0}=0$ on the 0-brane and
\eqn\bc{g_{ab}\partial_{\sigma}x^{b} + 2\pi \apm B_{ab}
\partial_{\tau}x^{b}|_{\sigma=\pi}=0}
on the 2-brane. These boundary conditions may be diagonalized by
working with the complex linear combination $X=x^1+ ix^2$.
\bc\ implies the shifted mode expansion \eqn\modeb{\eqalign{X= &
{i\over \sqrt{2}} \sum_{n=-\infty}^{\infty}[
e^{i(n+\nu)(\tau+\sigma)}-e^{i(n+\nu)(\tau-\sigma)}]
\frac{\alpha_{n+\nu}}{n+\nu}\cr \bar X = & {i\over \sqrt{2}}
\sum_{n=-\infty}^{\infty}[
e^{i(n-\nu)(\tau+\sigma)}-e^{i(n-\nu)(\tau-\sigma)}]
\frac{\bar\alpha_{n-\nu}}{(n-\nu)},}} where 
$(\a_{n+\n})^{\dagger}=\bar \a_{-n-\n}$,
\eqn\tw{e^{2 \pi i
\nu} = - \frac{1+ i b}{1-ib},\;\; \nu \in [0,1)} and $b
=\frac{ 2 \pi \apm B_{12}}{g}.$ In the limit \limlarge\ $b \gg 1$ , and
\tw\ simplifies to \eqn\lim{\nu = 1 - \frac{1}{\pi b}.} 

World-sheet superpartners of $x^1,x^2$ fields are similarly affected by
turning of $B$ field in those directions. The moding of fermions in the 
NS sector is in the usual way shifted by a half-integer
relative to that of the bosons. Puting
$\psi_{\pm}=(\psi^1+i\psi^{2})_{\pm},$ we have
\eqn\modef{\eqalign{\psi_{-}=&\frac{1}{\sqrt2}\sum_{n=-\infty}^{\infty}
e^{i(n+\nu - \frac{1}{2})(\tau-\sigma)}\psi_{n+\nu - \frac{1}{2}}\cr
\bar{\psi}_{-}=&\frac{1}{\sqrt{2}}\sum_{n=-\infty}^{\infty}
e^{i(n-\nu +\frac{1}{2})(\tau-\sigma)}\bar{\psi}_{n-\nu +\frac{1}{2}}}} 
where $\psi_{n +\n - \half}^{\dagger}=\bar \psi_{-n-\n +\half}$. 
$\psi_{+}$ is not independent, but is determined in terms of $\psi_{-}$
by the NS boundary condition. 

Upon
quantization, $\alpha$'s and $\psi$'s become creation and 
annihilation operators, whose 
non-vanishing (anti)commutators are
\eqn\cm{[\alpha_{m+\nu},\bar\alpha_{n-\nu}]=(m+\nu)\delta_{m+n}; ~~~~~
\{\psi_{m+\nu-1/2},\bar\psi_{n-\nu+1/2}\}=\delta_{m+n}.}

For completeness, we present expressions for the operators
$L_m, G_m$. 
Then 
\eqn\nnL{\eqalign{L_m =&\left(\sum_{n=-\infty}^{\infty}
\alpha_{m+n+\nu}\bar\alpha_{-n-\nu}+
(\frac{m-1}{2}+n+\nu)\psi_{m+n+\nu-{1\over 2}}
\bar\psi_{-n-(\nu-{1\over 2})}\right)\cr+ 
&{1\over 2}\left(\sum_{n=-\infty}^{\infty}\alpha_{m+n}\cdot\alpha_{-n}+
(\frac{m-1}{2}+n)\psi_{m+n-\half}\cdot \psi_{-n+\half}\right),}}
where  the last term runs over DD and NN oscilators, so for example
$\alpha_m\cdot\alpha_n = -(\alpha_t)_m(\alpha_t)_n +
(\alpha_i)_m\alpha^i_n$,
and \eqn\nonum{G_{m+\frac{1}{2}} =\sum_{n=-\infty}^{\infty}
\left(\alpha_{m+n+\nu}\bar\psi_{-n-(\nu-{1\over 2})}
+\psi_{m+n+\nu+{1\over 2}}\bar\alpha_{-n-\nu}+
 \alpha_{m+n}\psi_{-n+\half}\right).}

\listrefs
 \end